# Active bialkali photocathodes on free-standing graphene substrates

Hisato Yamaguchi[1,a)], Fangze Liu[1], Jeffrey DeFazio[2], Claudia W. Narvaez Villarrubia[1], Daniel Finkenstadt[3], Andrew Shabaev[4], Kevin L. Jensen[5], Vitaly Pavlenko[6], Michael Mehl[4], Sam Lambrakos[6], Gautam Gupta[1], Aditya D. Mohite[1,b)], Nathan A. Moody[6,c)]

[1]MPA-11 Materials Synthesis and Integrated Devices (MSID), Materials Physics and Applications Division, Mail Stop: K763, Los Alamos National Laboratory, P.O. Box 1663, Los Alamos, New Mexico 87545, U.S.A.
[2]PHOTONIS USA Pennsylvania Inc., 1000 New Holland Ave., Lancaster, PA 17601, U.S.A.
[3]Physics Department, U.S. Naval Academy, Stop 9c, 572c Holloway Rd., Annapolis, MD 21402, U.S.A.
[4] Center for Materials Physics and Technology (Code 6390), Materials Science and Technology Division, Naval Research Laboratory, Washington DC 20375, U.S.A.
[5] Materials and Systems Branch (Code 6360), Materials Science and Technology Division, Naval Research Laboratory, Washington DC 20375, U.S.A.
[6]Accelerators and Electrodynamics (AOT-AE), Accelerator Operations and Technology Division, Mail Stop: H851T, Los Alamos National Laboratory, P.O. Box 1663, Los Alamos, New Mexico 87545, U.S.A.

Authors to whom correspondence should be addressed. Electronic mail: [a)] Hisato Yamaguchi (hisatoyamaguchi08@gmail.com), [b)] Aditya D. Mohite (amohite@lanl.gov) , [c)] Nathan A. Moody (nmoody@lanl.gov)

## Abstract

The hexagonal structure of graphene gives rise to the property of gas impermeability, motivating its investigation for a new application: protection of semiconductor photocathodes in electron accelerators. These materials are extremely susceptible to degradation in efficiency through multiple mechanisms related to contamination from the local imperfect vacuum environment of the host photoinjector. Few-layer graphene has been predicted to permit a modified photoemission response of protected photocathode surfaces, and recent experiments of single-layer graphene on copper have begun to confirm these predictions for single crystal metallic photocathodes. Unlike metallic photoemitters, the integration of an ultra-thin graphene barrier film with conventional semiconductor photocathode growth processes is not straightforward. A first step toward addressing this challenge is the growth and characterization of technologically relevant, high quantum efficiency bialkali photocathodes grown on ultra-thin free-standing graphene substrates. Photocathode growth on free-standing graphene provides the opportunity to integrate these two materials and study their interaction. Specifically, spectral response features and photoemission stability of cathodes grown on graphene substrates are compared to those deposited on established substrates. In addition we observed an increase of work function for the graphene encapsulated bialkali photocathode surfaces, which is predicted by our calculations. The results provide a unique demonstration of bialkali photocathodes on free-standing substrates, and indicate promise towards our goal of fabricating high-performance graphene encapsulated photocathodes with enhanced lifetime for accelerator applications.

# Introduction

The evolving needs of advanced electron accelerator-based x-ray light sources, such as energy recovery linacs (XERL) and free electron lasers (XFEL), have motivated continual development in photocathode materials with a particular emphasis on high brightness and long lifetime[1-4]. Semiconductor materials, particularly those of the alkali-antimonide family, are of keen interest as their relatively high quantum efficiencies and low thermal emittances can yield a high brightness beam. Their characteristic sensitivity to the local gas environment, however, limits them to short operational lifetimes in most accelerator vacuum environments[1]. This sensitivity is typically due to ion or thermally induced damage at the surface, and can include changes in stoichiometry due to cesium loss, electrolysis, and/or material decomposition[5,6]. For the general considerations of vacuum sensitivity, studies of quantum efficiency degradation of the alkali-antimonides have found that the dominant effect is due to surface poisoning through irreversible reactions with residual reactive gases containing oxygen[2]. Given that graphene has been shown to be impermeable to gas molecules[3], even hydrogen and helium[4-6], a goal of this study is to examine the feasibility of utilizing graphene as an atomically flat, ultra-thin barrier to passivate the sensitive photocathode surface and prevent degradation in photosensitivity[7]. The ability to control the number of layers and grow relatively large area (e.g., ~1 $cm^2$) graphene thin-films using chemical vapor deposition (CVD) provides a path for investigating the protective barrier properties[8,9]. Additionally, limiting the thickness to a few atomic layers can preserve transmission of the photoexcited electrons into vacuum. A critical first step in investigating the barrier properties of graphene is the ability to mechanically and electrically interface the photocathode with graphene while not compromising the photocathode through thermal decomposition or contamination. Synthesis of graphene requires high temperature and pressure of precursor gases, while the synthesis of the bialkali photocathode ($K_2CsSb$) requires relatively lower temperatures and extreme vacuum purity. Bringing these compounds together required an innovative "top-down" approach, wherein ultra-thin few-layer graphene was separately grown and then suspended on a nickel-mesh scaffold. The bialkali cathode was then grown on the suspended graphene substrate in the open areas of the nickel mesh.

Using this suspended substrate method, we measured a quantum efficiency (the ratio of the number of emitted electrons $\Delta Q/q$ to the number of incident photons $\Delta E/\hbar\omega$, or $QE = (\hbar\omega/q)(\Delta Q/\Delta E)$) exceeding 5 % at the vacuum surface of the photocathode grown on few layer graphene, which demonstrates the potential to grow state-of-the-art alkali photocathodes with quantum efficiencies in the accepted range of 1 - 10% grown on well-characterized bulk substrates[1,10]. This strategy overcomes the complications that could arise using the conventional 'bottom-up' approach using a complicated *in-vacuo* dry transfer technique, which remains a challenge to a.) secure a ~mm range defect-free area suitable for photocathode applications, and b.) execute the transfer in pristine vacuum. Moreover, our approach demonstrates the feasibility of photocathode synthesis on free-standing ultra-thin substrates, which may be of use in adjacent applications. We observe an increase in effective work function of the photocathode at the graphene interface, and this is compared to theoretical calculation results. Validation of graphene-cathode compatibility, as evidenced by stable photoemission (from both the coated and un-coated sides) and correlated perturbation of spectral response, opens new avenues to cathode designers. In particular, it demonstrates, for the first time, the critical functions of charge injection, charge extraction, mechanical support, and optical transparency in the entirely new configuration of free-space suspension of cathode films on few monolayer substrates.

## Results and Discussion

### I. Overall concept and measurement setup

The design of free-standing graphene substrates on mesh grids as shown in Figure 1(a) provides a unique opportunity to probe photoresponse from both sides. This is critical for studying the effect of the graphene substrate on the photocathode performance, as will be discussed in subsection III. The schematic in Figure 1(a) depicts the overall concept, where a 2 μm thick nickel mesh grid with unit cell aperture of 6.5 μm and open area ~42μm$^2$ is used to suspend a graphene substrate with thickness between 5-8 monolayers. Figure 1(b) shows the arrangement for photocurrent measurement: a patterned anode is deposited on the inside of the windows which preserves transparency while facilitating electron collection; the excitation laser can be easily positioned within the patterned anode material to impinge upon the active cathode area at a desired location. Photocathode deposition occurs on the nickel (Ni) mesh side of the graphene substrate, such that the mesh would function as the charge injection electrode. For photocathode characterization, a vacuum-sealed package was devised with optically transparent sapphire windows on both sides of the photocathode. The sealed construction is similar to that of a commercial phototube and allows for long-term storage and study. Photoelectrons emitted from the desired surface are measured using a low-noise nanoammeter while a bias voltage is applied to charge-collecting anodes (Figure 1(b), see Methods section for details). Figure 1(c) shows an end-view of the setup with a red circle indicating a representative optical spot.

**II. Photocathode deposition on free-standing graphene substrates**

Graphene films of five and eight layer thickness (5L and 8L hereafter, respectively) were prepared as free-standing substrates to investigate the performance of bialkali antimonide photocathodes. Monolayer graphene was stacked *via* transfer process instead of direct multilayer growth to achieve high reproducibility and controlled thickness (see Methods section for details). The graphene used in this study was grown by chemical vapor deposition (CVD) method on copper (Cu) substrates. The graphene quality was tested using Raman spectroscopy, atomic force microscopy (AFM), and scanning electron microscopy (SEM), following a well-stablished wet-transfer process[11] on Si/SiO$_2$ substrates. Characterization results indicated that the transferred monolayer graphene had minimal crystal structure defects on the scale of ~1 μm, and thickness was ~0.5 nm. Specifically, Raman spectra for the film exhibited two dominant peaks of G and 2D, around ~1,580 cm$^{-1}$ and ~2,700 cm$^{-1}$, respectively, and absence of D peak around 1,350 cm$^{-1}$ (Figure 2(a)). The D peak is indicative of defects or discontinuities in the carbon network because the A$_{1g}$ zone-boundary Raman mode of carbon atoms requires their presence. An absence of this vibrational mode is indicative of high crystal quality. Raman spectra can also be used to identify the thickness of graphene using the ratio between the heights of the 2D and G peaks. The value for our graphene was 2D/G~3, which is well above an accepted characteristic value of 2 for monolayer graphene[12]. The thickness of the monolayer graphene was further confirmed by cross-sectional height profile using AFM, revealing a height of ~0.5 nm (Figure 2(b)). In addition, AFM studies confirmed a smooth and flat surface of monolayer graphene without observable pinholes or ruptures, which are consistent with the Raman results indicating minimal structural defects in this film (see Figure 2(c), where scale bar is 1 μm). The uniformity of monolayer graphene for broader regions was confirmed by scanning electron microscopy (SEM) (see Supplementary Information).

Nickel mesh transmission electron microscopy (TEM) grids with a mesh aperture of 6.5 μm were used as a supporting scaffold for the 5L and 8L graphene substrates. As previously mentioned, TEM grids instead of rigid substrates were used to allow for subsequent deposition of bialkali cathode material on graphene in a manner that would allow for photoemission from both the graphene-coated and uncoated sides. A wet-based process was used for transfer of graphene to the TEM grids[11] (see Methods section for details). Figure 2(d) shows optical microscope images of as-prepared 8L graphene substrates (scale bar is 100 μm), and optical microscope images of the TEM grid without graphene are shown in Supplementary Information for comparison. It is clear that very thin graphene sheets are able to span the mesh holes with 6.5 μm diameter without observable pinholes or ruptures. Reliably spanning the open area of a TEM mesh is itself a practical confirmation of the transfer process. Nearly 100 % coverage of the 2mm diameter TEM mesh grid was confirmed for 5L and 8L graphene substrates using optical microscopy (see Supplementary Information for details).

For the photocathode deposition, the TEM grid-supported graphene substrates were assembled onto a pre-cleaned stainless steel (SS) outer frame assembly in a 3 x 3 array structure (see Supplementary Information). The graphene substrates in the assembly were oriented such that the photocathode was deposited on the exposed Ni mesh side in order for the Ni mesh to act as the charge injection electrode. Photoemission current was monitored as a process metric during deposition using a SS witness substrate, and the stoichiometry of typical resulting photocathode films using this process has been previously confirmed as $K_2CsSb$[12,13]. The entire frame assembly was then sealed in a vacuum package using conventional phototube techniques. The specially designed vacuum package also integrates the aforementioned anode grid patterns on the sapphire windows. The grids and the cathode are all electrically floated with respect to the chassis of the tube and this arrangement minimizes leakage current between the photocathode and the anodes (see Supplementary Information). This type of vacuum-tube photocathode enclosure provides an extremely stable environment for long-term experiments (much longer than what is observed in a dynamically pumped vacuum environment).

**III. Photoemission measurements**

Validation of graphene-photocathode compatibility was performed by measuring the quantum efficiency (QE) as a function of wavelength (spectral response) of photocathodes deposited on 5L and 8L graphene substrates. Of the four possible combinations of incident light "in" and photoemission current "out" as indicated in Figure 1(a) (i.e., photocathode (PC)-in / graphene (Gr)-out, PC-in / PC-out, Gr-in / PC-out, Gr-in / Gr-out), the spectral response from the Gr-in / Cs-out configuration (transmission mode) was found to provide the most useful method for studying $K_2CsSb$ deposited on graphene substrates. The reason is that only minimal and reliable corrections are required to obtain absolute QE values (see Supplementary Information for details). Figure 3(a) shows the spectral response in this configuration for the photocathodes deposited on 5L and 8L graphene substrates, where persistent photoemission was achieved with peak QE of 5.69 % at 3.14 eV (395 nm) and the spectral response exhibited characteristic features similar to that of $K_2CsSb$ reference cathodes (indicated by the purple arrows). Graphene transmission loss of 2.3 % per monolayer is taken into an account. A QE value exceeding 5 % in the UV from $K_2CsSb$ photocathodes on free-standing few-layer graphene substrates provides a clear demonstration of photocathode-graphene substrate architecture as a promising high QE photocathode with enhanced lifetime and chemical robustness[1,10,13] as has been sought since the emergence of photocathodes as a bunched electron source for injectors[14]. It should also be noted that the peak in QE between 3-3.5eV evident in Figure 3(a) is a general characteristic shared among the family of the alkali antimonide photocathodes. The primary cause is an onset of electron-electron (*e-e*) collisions: *e-e* scattering events are prevented until the final states of both electrons can reside in the conduction band of $K_2CsSb$[15]. This means that for photon energies below the observed peak in QE, electrons excited from the top of the valence band, for example, do not have enough excess energy to scatter with another electron (because their final states would lie within the bandgap). This leaves phonon scattering as the only scattering mechanism and QE continues to rise with increasing incident photon energy until the final states of two collisional electrons can both lie in the conduction band. This region corresponds to 3.0-3.5eV for alkali antimonides, and QE drops even as incident photon energy increases (as seen in Figure 3(a)). Another contributing factor to the peak QE in the region of 3.0-3.5eV is that optical absorption is highest in this region (as demonstrated in Figure 3(b)), as is optical penetration depth. The resulting shape and fine structure of the spectral response curve is therefore unique to the particular film in question (e.g., with a particular stoichiometry, crystallinity, thickness). The nearly perfect match between the spectral response curves in Figure 3(a), corresponding to 5L and 8L graphene substrates, eliminates the possibility of graphene substrate thickness dependence on the intrinsic response of the deposited $K_2CsSb$ photocathodes. Recall that the method for preparing graphene substrates (see Supplementary Information) allows unambiguous differentiation between 5L and 8L. The similar behavior of $K_2CsSb$ photocathodes on 5L and 8L substrates is further supported by optical transmittance: both cases showed signature peaks (indicated by the purple arrows in Figure 3(b)) except for ~7 % higher values for the 5L substrate solely due to difference in graphene layers. These photocathodes are highly stable in the vacuum-sealed phototube environment indicating no long-term adverse reaction between the $K_2CsSb$ and graphene substrates. Similar to a reference photocathode region deposited on the SS frame assembly there was no observable decrease in QE over a 100 minute stability measurement during current extraction from the photocathodes on the graphene substrates (Figure 3(c)). Furthermore, no degradation was observed over a 6+ month period since the time of the photocathode fabrication.

Despite the demonstrated functionality of $K_2CsSb$ on graphene substrates, the resulting QE is lower at all wavelengths compared to that obtained using conventional bulk substrates. For instance, transmission-mode $K_2CsSb$ on glass or sapphire bulk substrates can exhibit peak QE in excess of 25% in the region of ~ 3.5 eV (Supplementary Information). Within the tube assembly a significantly higher QE was obtained from photocathode film on the stainless steel reference regions than on the graphene substrates. The suppressed spectral response could be caused by a number of factors. A likely hypothesis is an initial presence of residual water molecules on the graphene substrates. This is unlikely for conventional bulk substrates given the extended high-vacuum bakeouts utilized in industrial photocathode production, however in the case of few-layer graphene substrates a trace amount of adsorbed water molecules could remain trapped in between the layers during their wet-based stacking processes and present a local source of contamination to the photocathodes. This could be particularly true in this study due to the fact that the suspended graphene was not subjected to a prolonged anneal. The graphene substrates were thermally annealed at only 100 °C in air after each of the transfer steps to minimize trapped water at the interlayer. However, an in-vacuum anneal is required to eliminate the possibility of residual water and this study is planned for the near-future. Earlier investigations have shown that residual water can persist even during thermal annealing at 350 °C in UHV, due to their geometrical capping by the graphene basal planes[16-18]. From optical transmission measurements the $K_2CsSb$ film thickness on graphene substrates were estimated to be similar to that obtained on conventional substrates under similar growth conditions (> 20 nm), suggesting that differences in adhesion coefficients and the resulting film thicknesses play a minimal role in the observed QE difference over conventional substrates. Spectral features in optical transmission of the $K_2CsSb$ films were observed to be less distinct on graphene compared to that of reference deposited on sapphire substrates (Supplementary Information). This maybe an indication of $K_2CsSb$ film degradation due to the presence of the graphene substrate, or *vice versa*, and details are to be investigated in a follow-on study. Regardless, the deposited photocathode exhibited QE that exceeds 5 %, demonstrating that the critical function of charge injection, charge extraction, and mechanical support were satisfied in entirely new configurations of free-space suspension of bialkali photocathode films on few monolayer substrates.

The validation of graphene-photocathode compatibility and functionality was further investigated by studying the photocurrent emitted on the graphene side. The measurements provide insight on the photocathode performance at the graphene interface in contrast to the photoemission obtained from the non-graphene side. It is worthwhile noting again that graphene is one of the very few material options available for this purpose, if not the only one: it satisfies all requirements of being optically transparent and providing a mechanically stable support while prior studies of monolayer graphene on planar Cu photocathodes have shown that the graphene barrier does not significantly block electron transport (tunneling) through it as part of the so-called "three-step" photoemission process. For this study of Gr-side electron emission, the cathode was illuminated from the back (Gr) side (illumination from the $K_2CsSb$ (PC) side could also provide information about photocathode performance at the graphene interface albeit in a less direct manner). Photoemission was indeed observed from the graphene side, and including the necessary optical corrections (Supplementary Information) the QE was measured at ~ 1.0 % at 4.5 eV, significantly lower than that for direct photoemission into vacuum (i.e. emitted to the PC side) for the same specimen as indicated in Figure 4(a). Furthermore drastic changes in the spectral response features were observed as shown in Figure 4(b), suggesting a change in the electronic structure of $K_2CsSb$ at its interface with graphene (to be discussed in comparison with our calculation results). To gain insight into the change that would impact the photocathode performance, we examined the work function of this $K_2CsSb$-graphene system. The effective work function can be determined by observing the low-energy cutoff of Figure 4(b). Specifically, this data was re-plotted using the square root of QE for the vertical axis, and the linear region of the data was extrapolated to the horizontal axis as shown in Figure 4(c)[2]. The results indicate that the work function increased by ~0.3 eV (Figure 4(d)) compared to the direct photoemission (PC) side, where the cut-off energy was consistent at ~1.8 eV regardless of substrate thickness. These results suggest that while there appears to be a drastic change of $K_2CsSb$ electronic structure at the graphene interface, it does not affect the electronic structure on the vacuum-interfacing side of the photocathode film, which has estimated thickness of > 20 nm. Considering the electron escape depth for the energy range of incident photons, the thickness of $K_2CsSb$ with intrinsic electronic structure can be estimated to be more than ~10 nm[19].

Importantly, the lower QE observed from Gr-side photoemission (compared to that of PC-side) agrees with theory prediction, but the cause of apparently negligible difference in QE between the 5L and 8L graphene substrates remains unclear. If the collected photoelectrons are originating from $K_2CsSb$ *via* tunneling through graphene, the expected result would be a strong dependence on graphene thickness correlated to QE. Quantum tunneling introduces an exponential decay in the transmitted current that scales with the product of the barrier width with the square root of the barrier height above the electron energy level[20] (there is ~1.5 nm thickness difference between the thickness of 5L and 8L). However, the Gr-side QE for $K_2CsSb$ had the same order of magnitude for both 5L and 8L, suggesting that a simplistic notion of tunneling through graphene is not a sufficient description. An alternate explanation of the similarity in response for the case of 5L and 8L is the possibility of diffusion of element(s) through free-standing graphene substrates to the non-deposited side during photocathode deposition, which we observed for a case of $Cs_3Sb$ (Supplementary Information). Cs is an element with known high diffusivity thus it is useful to consider the possibility that the Gr-side photosensitivity originates from photocathode material which diffused onto the graphene layer during fabrication. Note, however, that the shape of the spectral response curve for photoemission from the graphene side does not resemble that of the reference sample or that from the opposite side. Thus, if photosensitivity is due to diffusion, the diffused material does not share the same stoichiometry as $K_2CsSb$. Possible diffusion paths of Cs could be either at macroscopic in-plane structural defects of graphene or narrow cracks at the edges of photocathode films. A counterpoint to this explanation, however, is that the resulting layer would likely be very thin and not able to account for the comparatively large photocurrent observed. Although the interface between graphene and diffused photocathode material could still provide information on electronic structure, the limited amount of diffused photosensitive material on the non-deposited side of graphene may be the origin of the observed Gr-side photoemission. In the planned follow-on study, graphene substrates will be prepared with much fewer macro-scale structural defects (>5 μm) (which may be induced during the transfer process such that permeation of alkali material would be significantly reduced or eliminated entirely).

**IV. DFT calculations for band diagram at the interface**

An increase in work function for $K_2CsSb$ by the presence of interfacing graphene is indeed found in our calculation results, as discussed below. To perform calculations significant expansion of the unit cell for graphene was required, namely its 7x2 reconstruction and appropriate resizing of the $K_2CsSb$ (111) unit cell to match their lattices. A Cs-rich surface was chosen beneath graphene, as recommended by experiment, and graphene was placed symmetrically on both sides of the slab resulting in a total cell stoichiometry of $(K_8Cs_7Sb_5C_{28})_2$. This reflects 13 layers of $K_2CsSb$ slab between two layers of graphene, and a vacuum layer of at least 10 Å. The sandwiched structure was required simply for the calculation procedure purpose, and Figure 5(a) shows the unit cell for graphene on $K_2CsSb$ (111) from two views that correspond to the experiments described herein. To calculate the work function, the Fermi energy is subtracted in each case from the value of the electrostatic potential in the vacuum region. The electrostatic potentials, or rather pseudopotentials (from the VASP distribution), are depicted in Figure 5(b). The energy zeroes and vacuum electrostatic potential differ in the calculations depending on whether graphene is present or not, but energy differences, such as the work function, are well-defined and give the values of 1.75 eV and 3.70 eV by interfacing it with and without graphene, respectively. Notably, graphene presents a delta function-like well potential that scatters states transmitting from the bulk in a similar fashion as for potential barrier tunneling. A working theory for the change in work function and/or Fermi level is that electronegative adsorbates tend to increase the work function, although there are exceptions to this rule[21]. Adsorbates that are more electronegative than the substrate cause electrons to transfer to the adsorbate layer, causing an excess of negative charges on the outside and an excess of positive charges on the inside of the surface. This leads to a negative dipole pointing inward that reinforces the original surface dipole, causing the work function to increase. Cs is the most electropositive element thus we can expect electron transfer to the graphene, causing the work function to increase. The directions of dipole at the $K_2CsSb$-Graphene interface based on above mentioned explanation is consistent with our calculation results as shown in Figure 5(c).

The significance of these calculations is as follows. First, there is a charge transfer between the bulk $K_2CsSb$ and the graphene layer, indicating that the graphene layer is n-doped. Second, there is a change in the Fermi level of the graphene layer that translates to a change in the work function level of the graphene as a consequence of n-doping by the bulk $K_2CsSb$. This results in a reduction of work function for a single layer of graphene from 4.5 eV to 3.7 eV[22], that is higher than the work function of 1.75 eV for the bulk $K_2CsSb$ (see Figure 5(d)); this will partially account for the changes in work function observed experimentally. Third, as evident from Figure 5(b), short range features exist in the potential that affect transport, the details of which will be studied separately. Figure 4(c) shows how the interactions described above (between graphene and the adjacent $K_2CsSb$ layer) effect photoemission. Namely, emission from the PC-side of graphene shows no difference in QE and work function between the 5L and 8L cases, suggesting that for PC-side emission the charge transfer at the interface does not play a role because the electronic structure of $K_2CsSb$ remains the same far from the interface. It is reasonable to assume this conclusion is valid for arbitrary layer thickness, so long as mechanical integrity is preserved. On the opposite Gr-side, however, the interface *does* play a role and so work function increases compared to that of native $K_2CsSb$ and then it further increases from 5L to 8L. This trend of increased work function with increased layer thickness should be expected to continue for thicker graphene substrates. Substrates having fewer that 5L will be investigated in the future.

In general, photocathodes are sought which operate with the maximum possible wavelength permissible within total charge per bunch and emittance constraints. Unfortunately, these tend to oppose each other (both QE and emittance generally increase as wavelength decreases[23]). Particular wavelengths depend on the nature of the drive laser used, whether higher harmonics are generated through frequency doubling crystals, and how the graphene layers affect emission. A reasonable expectation, therefore, is that some optimization would be possible with regard to the wavelength chosen, and such questions will be part of the follow-on study.

## V. Summary Discussion

A technologically relevant, active photocathode film ($K_2CsSb$) has been grown and characterized on a novel substrate: few-layer graphene. This was accomplished in an encapsulated package which allows for long term storage and characterization. The changes in electronic structure were modeled and compared to experimental results. While there is an observed suppression in QE for films grown on 5 or 8 layer graphene, compared to films in this study which were grown on metallic substrates, the response is still well within the accepted 'active' range reported in the literature for alkali-antimonide cathodes grown on conventional substrates, see Fig. 1 in Ref[1]. A vanishingly thin substrate, such as graphene, can be expected to introduce multiple effects ranging from mechanical strain to compromised conductivity and charge injection into the film to potential trace contamination of water. These possibilities have been discussed in light of planned follow-on studies. The likely scenario of water molecules being introduced to the $K_2CsSb$ film will be studied further: anneal of the graphene substrates and its fixturing will occur prior to growth deposition for comparison. The results herein demonstrate a photocathode grown on a novel substrate with immediate and practical utility because its QE as a function of wavelength falls within the accepted ranges for the family of alkali antimonide emitters. The growth of $K_2CsSb$ on a novel substrate is the principal emphasis of this letter and it demonstrates significant potential for new technical approaches, such as gas barrier encapsulation, electron energy filtering, and transmission mode photocathodes, that were not otherwise possible and which could address long-standing challenges.

## Methods

### A. Graphene synthesis and characterization

Cu foils with size of 25 mm x 75 mm were placed in a quartz tube furnace and pre-cleaned under hydrogen gas flow at 1,000 °C for 30 minutes. Methane gas was then introduced into the tube for 15 minutes for graphene growth. After being cooled down to room temperature under vacuum, Graphene was transferred onto $SiO_2$/Si substrates (using the aforementioned wet-transfer technique) for material characterization by Raman spectroscopy and AFM (Figure 2). A WiTec

Alpha 300R spectrometer with 532 nm excitation was used for Raman spectroscopy. The laser spot size was less than 1 μm, and the power was reduced to prevent damage to graphene during the measurements. An atomic force microscope (Veeco Enviroscope) operated in tapping mode with standard cantilevers (tip curvature <10 nm) and spring constant of 40 N/m was used for AFM analysis.

**B. Preparation of free-standing graphene substrates**

Poly(methyl methacrylate) (PMMA) was spin coated on the top side of graphene-grown foils at 4,000 rpm for 60 seconds (MicroChem 495PMMA C4). After pre-bake of the PMMA-support at 180 ºC for 3 minutes, the entire specimen was immersed into Cu etchant of $CuSO_4$ and HCl aqueous solution for 12 hours. After the Cu foils were completely etched away, PMMA-supported graphene was rinsed in a water bath three times, and sized into 3 mm x 3 mm pieces for transfers. The monolayer graphene was then transferred onto $SiO_2$/Si substrates for stacking. The transfer process was repeated 5 or 8 times (for 5L & 8L, respectively). After each transfer, the PMMA-support was removed by immersing the entire specimen into an acetone bath heated to 60 ºC. A fresh acetone bath was prepared after 15 minutes and repeated three times, followed by an alcohol (IPA) bath at 60 ºC for 15 minutes. The PMMA-removed monolayer graphene was dried using an air blower and heated to 100 ºC for 5 minutes after each transfer to minimize residual water adsorption. Stacks of 5L or 8L graphene were then coated with PMMA-support for a final transfer onto 2,000 line per inch (lpi) transmission electron microscopy (TEM) Ni mesh grids (Ted Pella Inc. G2000HAN, 3 mm diameter, 2μm thickness, hole width 6.5 μm, open area 41%) using the procedures mentioned above. Intermediate transfer onto rigid substrates for a stacking was crucial for increasing the transfer yield. PMMA-supports on TEM mesh grids were removed using the above mentioned procedures of three acetone baths, one IPA bath, and hot-plate annealing at 100ºC.

**C. Bialkali antimonide photocathode deposition**

Graphene substrates on TEM Ni mesh grids were assembled into SS 3 x 3 array frames for bialkali antimonide photocathode deposition and vacuum-tube sealing at PHOTONIS USA. Thin SS foils with hole sizes of 1.5 mm diameter were used to retain the 2 mm diameter TEM grid supports, while some of the 3 x 3 array frames were left blank to allow reference photocathodes deposited on SS. The outer frames were electropolished and pre-cleaned along with the foils including a 600 °C bake in high vacuum. The assembled frames with graphene substrates and all vacuum envelope components were baked at 350 °C in UHV prior to *in-situ* photocathode deposition. The photocathode components K, Cs, and Sb were deposited on the SS and graphene substrates *via* thermal evaporation to achieve typical stoichiometry of $K_2CsSb$ with thickness of ~25-30 nm generally achieved on the SS. The vacuum package consisted of sapphire windows on both sides of the photocathode assembly with metal anode traces patterned on the inside of the windows to establish the necessary electric field for photoemission collection and measurement.

**D. Photoemission measurements**

A Newport 150W xenon (Xe) lamp with Oriel CS130 monochromator was used as a light source for photoemission measurements. Optical lenses were used to focus the spot size of incident light to ~1.5 mm, which is compatible with the graphene-spanning regions of the TEM support grids. The incident light was precisely aligned to the desired position in the 3x3 array to prevent stray light from impinging upon neighboring photosensitive regions.

Both anode patterns were biased with respect to the photocathode assembly to extract photocurrent on each respective side, where the photocurrent was collected and measured. It was confirmed that an applied bias of 90 V was enough to overcome space-charge effects and collect photoelectrons in this experimental setup (see Supplementary Information for details). A 1 MΩ resistor was connected to the SS outer frame support of the photocathode-deposited graphene substrates using a shielded coaxial cable, and a precision voltmeter was used to measure the voltage across the resistor (i.e., pico-ammeter) to obtain the photoemission current. The quantum efficiency was calculated using the known power of incident light at each wavelength, as obtained from a calibrated reference diode**.** The energy of the incident light was scanned from 2 to 4.5 eV while the corresponding photocurrent was recorded to obtain spectral response of the photocathodes.

**E. DFT calculations**

DFT calculations were performed considering an ideal system for the purposes of examining certain qualitative features of the graphene-$K_2CsSb$ interface. The ideal system, in contrast to the experimental configuration, was modeled as a symmetric arrangement of two layers of graphene around a layer of $K_2CsSb$ for computational feasibility. Electronic structure calculations were done using the Vienna Ab initio Simulation Package (VASP)[24-26], including core state effects *via* the VASP implementation[27] of Projector Augmented-Wave (PAW) methods[28]. We used the Local Density Approximation (LDA)[29,30] to Density Functional Theory (DFT)[31,32] which is the most basic functional available in VASP. In DFT packages, there are a myriad of functional choices; particularly, geometry/lattice-constants present large differences between generalized-gradient, van der Waals-corrected and other hybrid functional calculations, compared to the LDA. This has been discussed in previous studies on graphene[33], as well as on other metallic[34,35] and dielectric systems[36]. In the present study, however, LDA provides the superior method for calculating work-function, at least for comparisons to experimental values for elemental surfaces[37]. LDA also provides a correct graphite interlayer spacing and surface bonding distance. For this reason, we consider herein only LDA functionals. We used the VASP-supplied LDA potentials for C, Cs, K, Sb (specifically, Sep 2000, Mar 1998, Feb 1998 and May 2000 potentials, respectively) from the VASP distribution. All calculations use a kinetic energy cutoff of 500 eV, which is 20% larger than the suggested cutoff for Carbon (400 eV). Generally the cutoff might be chosen higher for a better converged surface energy, but the work-function itself was seen to depend very little on cutoff value. For all structures, a Monkhorst-Pack k-point grid[38] with spacing between 0.2 and 0.3 per Å was chosen, adjusted to ensure a *k*-mesh consistent with the original unit cell symmetry. As an example, applying these settings, graphene in its primitive cell would use a 15x15x1 mesh. The *k*-mesh is forced to be centered on the gamma point. The MedeA software system[39] was used to drive VASP in calculations of the optical properties, electrostatic potential and some other calculations.

## Supplementary Information

Supplementary Information is available from the *npj 2D Materials and Applications* website (http://www.nature.com/npj2dmaterials/).

## Acknowledgments

Authors acknowledge financial supports from the Los Alamos National Laboratory (LANL) Laboratory Directed Research and Development (LDRD) Program through Directed Research (DR) "Applied Cathode Enhancement and Robustness Technologies (ACERT)" (Project #20150394DR). Studies were performed, in part, at the Center for Integrated Nanotechnologies, an Office of Science User Facility operated for the U.S. Department of Energy (DOE) Office of Science. LANL, an affirmative action equal opportunity employer, is operated by Los Alamos National Security, LLC, for the National Nuclear Security Administration of the U.S. Department of Energy under contract DE-AC52-06NA25396. Authors also acknowledge Charudatta Galande of Rice University and Akhilesh Singh of LANL for their experimental supports on CVD graphene growth.

**Contributions**

N.A.M. conceived the project. H.Y. organized the project with supports from N.A.M. H.Y., N.A.M., and A.D.M. designed the experiments. H.Y. prepared graphene substrates. J.D. fabricated bialkali photocathodes on graphene and sealed them in a vacuum tube. F.L. performed photoemission measurements of photocathodes and atomic force microscopy on graphene. C.N.V. performed Raman spectroscopy on graphene and arranged XPS measurements. D.F. performed DFT calculations with inputs from A.S., K.J., M.M., and S.L. F.L. and V.P. deposited bialkali photocathodes on graphene for material characterization. A.D.M. and G.G. oversaw the graphene related efforts. H.Y. wrote the manuscript with supports from N.A.M. and inputs from all authors.

**Figure Captions**

**Figure 1** (a) Concept of graphene substrate photocathode (b) Schematic of experimental setup (c) Photograph of graphene substrate photocathodes sealed in a vacuum tube. Red circle indicates an optical spot size used for the measurements.

**Figure 2** Graphene substrate characterizations by (a) Raman spectroscopy (b) AFM height profile (c) AFM topology (d) Optical microscopy. Height profile in (b) is along the dashed line in (c). Scale bars in (c) is 1 μm, (d) is 100 μm.

**Figure 3** (a) QE spectral response (b) Optical transmission (c) Stability of $K_2CsSb$ photocathodes deposited on graphene substrates. For (a) and (b), red circles and black squares are for the photocathodes on 5 and 8 layer graphene substrates, respectively. Purple arrows indicate the regions with signature features. Black line in (c) is for a photocathode on SS substrate as a reference, and purple line is for photocathode on 8 layer graphene substrate. QE is normalized in (c) to the photocathode on SS substrate.

**Figure 4** QE spectral response based on photoemission from the photocathode (PC) and graphene (Gr) sides of the films. GrPC indicate photoemission measured in a Gr-in, PC-out configuration and GrGr in a Gr-in, Gr-out configuration. (a) QE in % (b) QE normalized to the highest value from the PC side (c) Photon energy cutoff indicating the work function. Extrapolating lines are shown in dash. (d) Work function of the photocathodes deposited on the graphene substrate with different thicknesses. Gr thickness = 0 is for photocathode deposited on SS as a reference.

**Figure 5** (a) Unit cell of graphene on $K_2CsSb$, with C, Cs, K, Sb atoms shown as black, blue, red and yellow, respectively. Views are shown along [111] direction (left) and (111) planes (right). (b) The local electrostatic potential without (left) and with (right) graphene on the surface of $K_2CsSb$. All curves are averaged over the directions transverse to the surface normal. The red curve shows a suitable macroscopic average over approximately 5 Å to demonstrate the energy zero level, relative to vacuum. Inset is enlarged (c) Direction of dipole at graphene-$K_2CsSb$ interface. Blue and red regions indicate electrons are being lost and gained, respectively.

Specifically, electrons are being lost from the Cs atoms (regions indicated by black circular lines), and are gained by the C atoms (red honeycomb regions). (d) Band diagram at graphene-$K_2CsSb$ interface with inputs from (b) and (c).

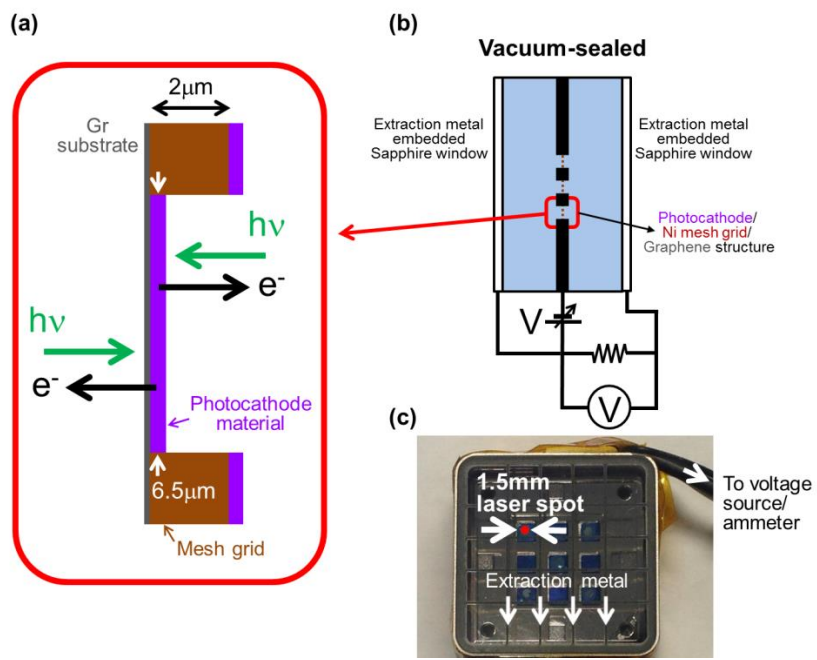

Figure 1

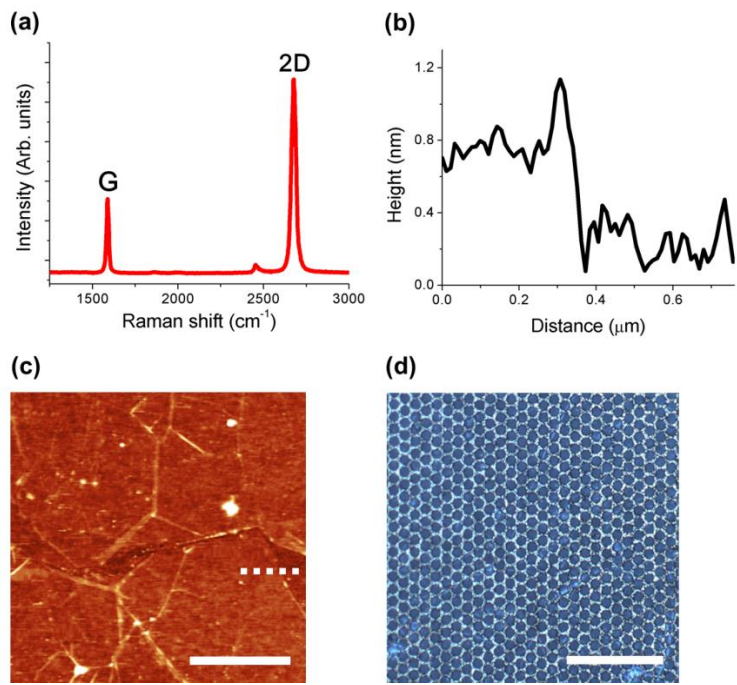

Figure 2

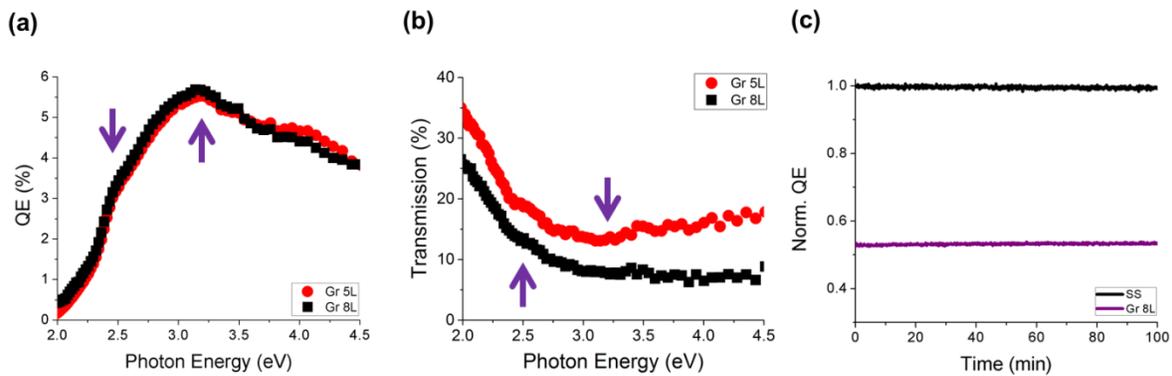

Figure 3

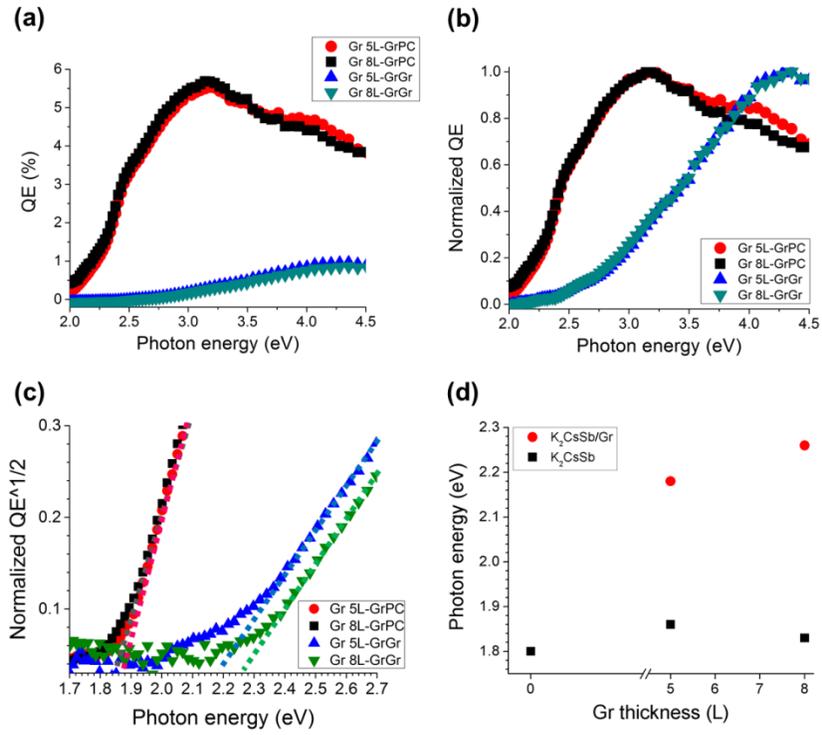

Figure 4

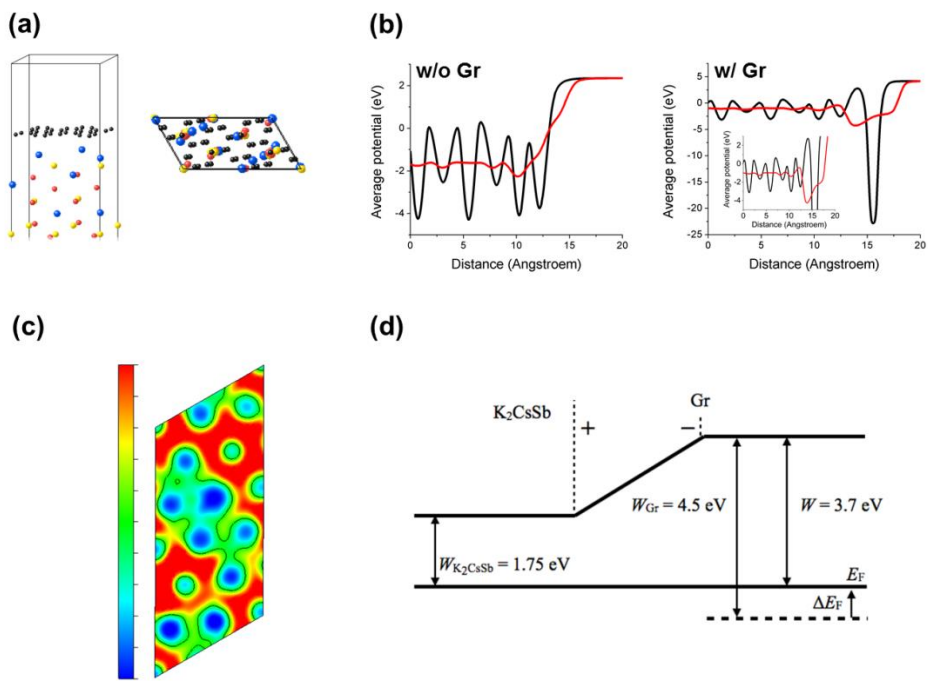

Figure 5

Supplementary Information for
## Active bialkali photocathodes on free-standing graphene substrates

Hisato Yamaguchi[1,*], Fangze Liu[1], Jeffrey DeFazio[2], Claudia W. Narvaez Villarrubia[1], Daniel Finkenstadt[3], Andrew Shabaev[4], Kevin Jensen[5], Vitaly Pavlenko[6], Michael Mehl[4], Sam Lambrakos[4], Gautam Gupta[1], Aditya D. Mohite[1,*], Nathan A. Moody[6,*]

[1]MPA-11 Materials Synthesis and Integrated Devices (MSID), Materials Physics and Applications Division, Mail Stop: K763, Los Alamos National Laboratory, P.O. Box 1663, Los Alamos, New Mexico 87545, U.S.A.
[2]PHOTONIS USA Pennsylvania Inc., 1000 New Holland Ave., Lancaster, PA 17601, U.S.A.
[3]Physics Department, U.S. Naval Academy, Stop 9c, 572c Holloway Rd., Annapolis, MD 21402, U.S.A.
[4] Center for Materials Physics and Technology (Code 6390), Materials Science and Technology Division, Naval Research Laboratory, Washington DC 20375, U.S.A.
[5] Materials and Systems Branch (Code 6360), Materials Science and Technology Division, Naval Research Laboratory, Washington DC 20375, U.S.A.
[6]Accelerators and Electrodynamics (AOT-AE), Accelerator Operations and Technology Division, Mail Stop: H851T, Los Alamos National Laboratory, P.O. Box 1663, Los Alamos, New Mexico 87545, U.S.A.

## Contents

*S1: Scanning electron microscopy (SEM) image of monolayer graphene*

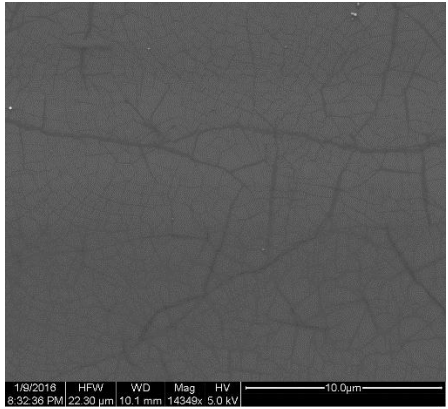

Figure S1 - SEM image of typical CVD monolayer graphene after transferred onto $SiO_2$/Si substrate using wet-based method.

*S2: Optical microscopy (OM) images of reference Ni TEM mesh grid*

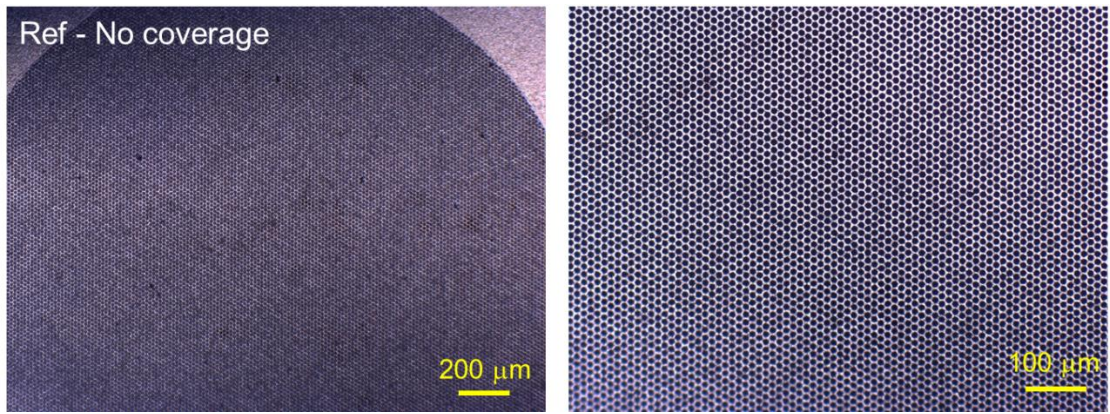

Figure S2 - OM images of reference Ni TEM mesh grid (no graphene).

*S3: Overall OM images of graphene substrates on Ni TEM mesh grid*

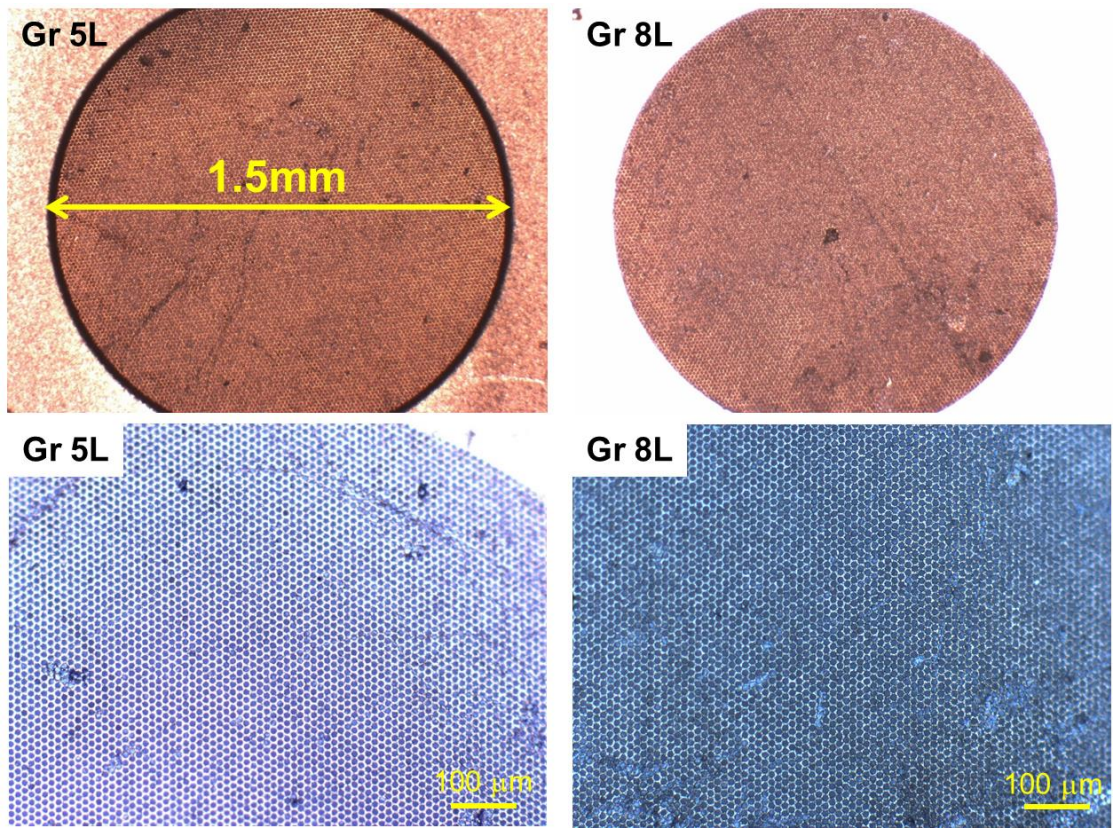

Figure S3 - Graphene substrates covering the entire Ni TEM mesh grid.

*S4: Photograph of graphene substrates assembled into SS frame*

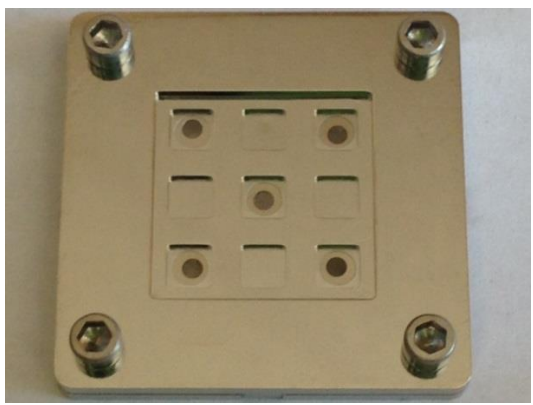

Figure S4 - Graphene substrates on Ni TEM mesh grids assembled into stainless steel (SS) mask foils with 1.5 mm diameter holes, sandwiched between SS outer frame with 3 x 3 array structure.

*S5: Photograph of vacuum-sealed phototube with electrical wire connections*

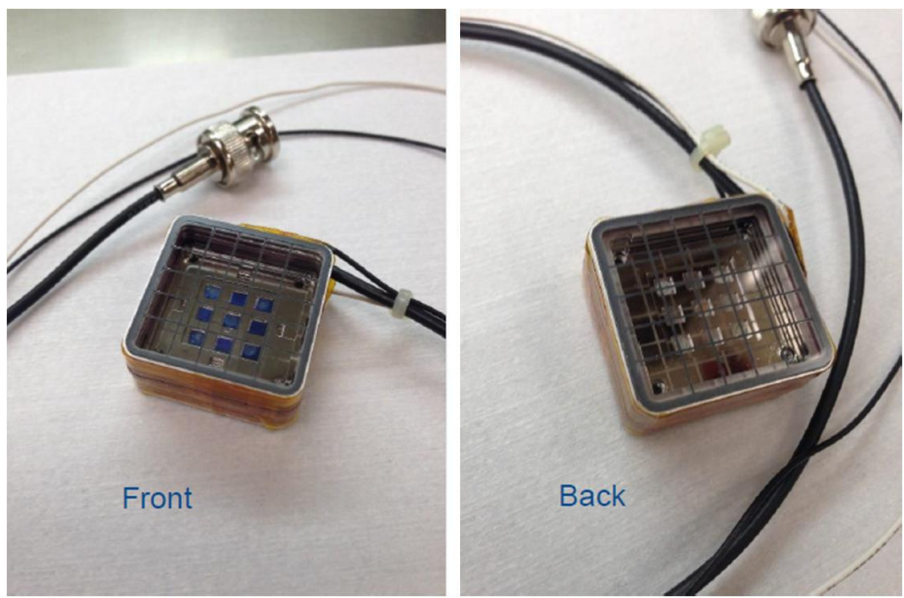

Figure S5 - Graphene substrates on Ni TEM mesh grids assembled into vacuum-sealed phototube. Photocathodes on graphene substrates and anode patterns on the sapphire windows are electrically connected to external wires. "Front" indicates the photocathode-deposition side, and "Back" indicates the graphene substrate side.

*S6: QE correction factor details*

There are three factors impacting the optical intensity to be considered when determining QE in this study. Specifically,

(1) Ni TEM mesh grid opening area: 2.44 (100% / 41%)
(2) Graphene transmission loss: 1.13 for 5L (100% / 88.5%), 1.226 for 8L (100% / 81.6%)
(3) Sapphire window transmission loss: 1.143 (100% / 87.5%)

The mesh correction factor (1) is based on the 41% open area of Ted Pella Inc. G2000HAN Ni TEM mesh grids. The graphene correction factor (2) is based on the well-accepted value of 2.3% optical transmission loss per graphene monolayer for 1-3 eV (Figure S6 (a)). The window loss factor (3) is due to reflections at the interfaces between the sapphire windows and air/vacuum. This transmission loss is slowly varying in the photon energy range used in this study as shown in Figure S6(b) (the sapphire windows used in the phototube were annealed at >1,400 °C, resulting in higher transmission than typically found in literature values). The raw data has therefore been multiplied by the product of these three factors yielding an overall correction of 3.15x (3.42x) the raw data for 5L (8L) respectively.

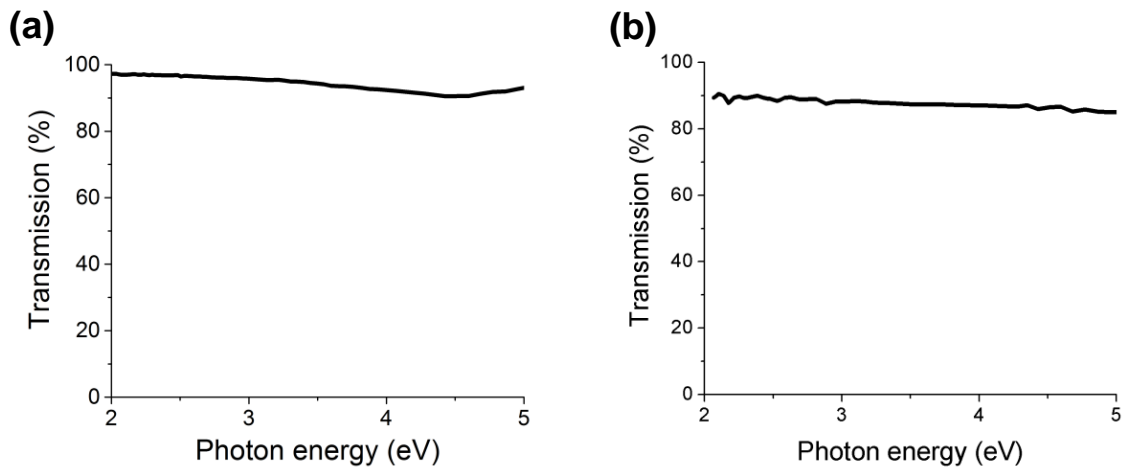

Figure S6 - Optical transmission of (a) monolayer CVD graphene and (b) sapphire windows used in the phototube.

We ensured that the focused light spot was within the 1.5 mm diameter SS mask holes thus no correction factor was necessary for incident photons being lost due to overfilling the SS mask region. The QE obtained for $K_2CsSb$ on graphene substrates was typically based on the transmission mode configuration (Gr-in, PC-out), as graphene absorption is the only specimen-related correction necessary to achieve QE in this configuration, and the graphene's optical absorption is very well-defined. The reflective mode configuration (PC-in, PC-out) may be less suitable for obtaining accurate QE since one would need the QE for $K_2CsSb$ on the Ni TEM gird substrates, which to date is not determined. Information extracted from photoemission on the Gr-out side is discussed in the main manuscript.

## S7: QE spectral response & optical transmission of K₂CsSb on reference sapphire substrate

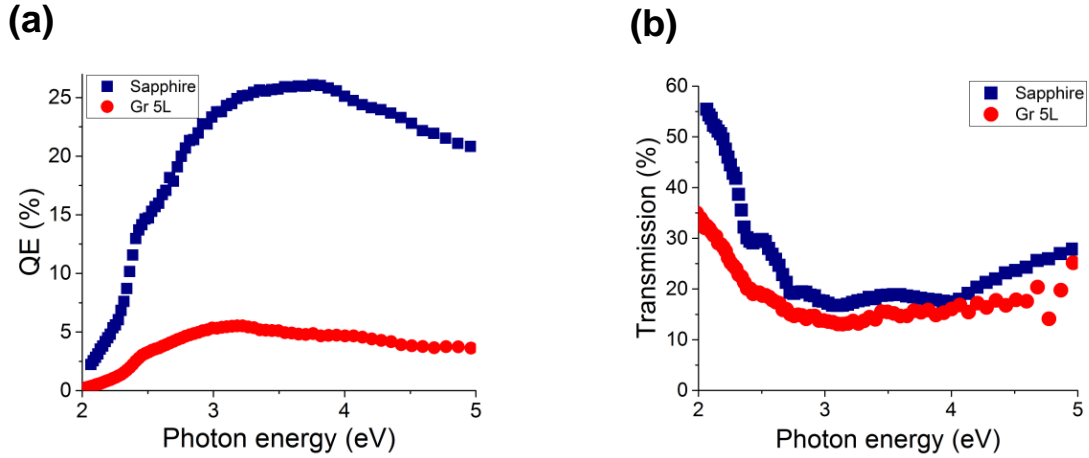

Figure S7 - (a) QE spectral response & (b) optical transmission of a typical $K_2CsSb$ on bulk sapphire ($Al_2O_3$) substrate in transmission mode (dark blue), corrected for optical loss at the air/sapphire interface. Response & transmittance of $K_2CsSb$ on a 5L graphene substrate measured in transmission mode is shown for comparison (red). Data is corrected as explained in S6.

## S8: SEM images of graphene substrates after Cs₃Sb photocathode deposition

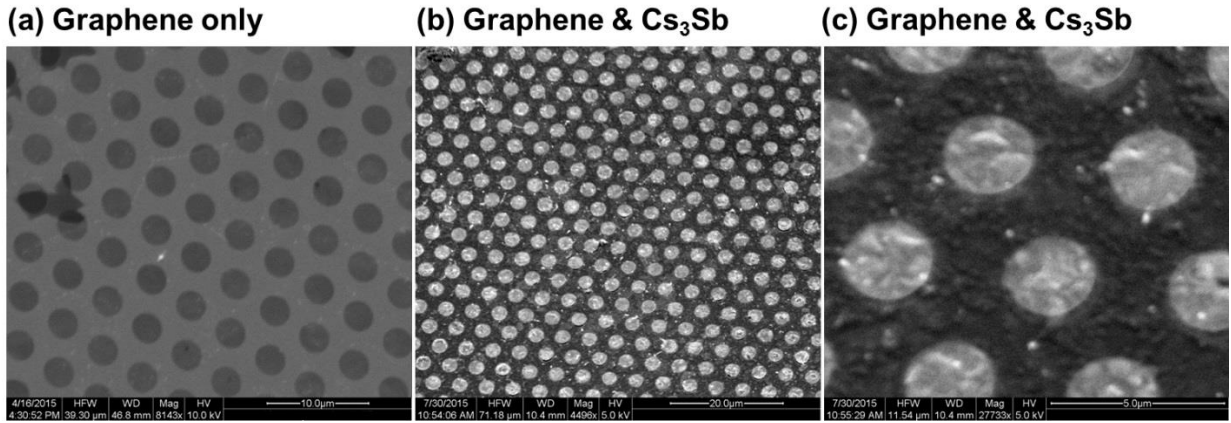

Figure S8 - SEM images of (a) free-standing graphene substrate on Ni TEM mesh grid prior to $Cs_3Sb$ deposition, and (b, c) after the deposition. All images are from graphene sides (non-deposited side).

## S9: X-ray Photoelectron Spectroscopy (XPS) of graphene substrates after Cs₃Sb photocathode deposition

XPS revealed that photocathode elements Cs and Sb are present on non-deposited side of the graphene-photocathode $Cs_3Sb$ structure (Fig. S9 (a)). The results suggest that the elements have diffused to non-deposited side of graphene substrates during the photocathode deposition. Our analyzing depth was kept shallow to below 2.2 nm by the use of 30º glazing angle. Given the graphene substrate thickness of at least 3 nm for individually stacked 5 layers, analyzing region did not overlap with the deposited side of graphene-photocathode structure. 3.7 atomic % of Cs and 0.4 atomic % of Sb was on the non-deposited side of graphene substrates in contrast to 9.9 atomic % Cs and 2.6 atomic % Sb on the deposited side. XPS on graphene substrate without photocathode deposition as a control had no detectable Cs nor Sb. Higher ratio of 9.3 for Cs to Sb on non-deposited side compared to the ratio of 3.8 for deposited side is consistent with higher diffusivity of Cs compared to Sb. This result strongly suggests that the presence of elements on non-deposited side involves diffusive process. Diffusion paths are not conclusive at this point; they could either be through nanometer pores of graphene sheets created during the synthesis process as natural defects or micrometer to even larger pores / cracks created during the graphene transfer or photocathode fabrication process. We are in the process of preparing photocathode-graphene structure with graphene that undergoes only one transfer as opposed to more than 5 transfers used in this study. The results will be reported elsewhere and they should provide insights as to which of above mentioned paths is dominant one.

Suzuki *et al.*[1] performed thorough investigation on residual water molecules on CVD graphene transferred *via* wet-based technique. By means of X-ray photoelectron spectroscopy (XPS), the presence of water molecules encapsulated in between graphene and substrate was strongly suggested by the oxygen (O1s) peak that did not diminish even at annealing temperature of 600 ºC. The peak is not observable for CVD graphene without a transfer. They performed Raman spectroscopy / electrical measurements on transferred graphene as complementary characterization and results were consistent with that of XPS. The presence of residual water molecules was further confirmed by rinsing the transferred CVD graphene in isopropyl alcohol (IPA) to remove them right after a transfer. Raman spectroscopy results indicated that IPA rinsing is indeed effective in removal of residual water molecules therefore structural defects in graphene induced by reaction between residual water molecules was clearly reduced for annealing temperature up to 600 ºC. We performed XPS on graphene transferred onto Ni mesh grids using wet-based technique and confirmed that similar amount of oxygen (11.7 atomic %) was present prior to annealing, which indicates a presence of residual water molecules (Fig. S9 (b)).

Severe decrease in quantum efficiency (QE) of bialkali photocathode by a trace amount of moisture is well-established knowledge in industry as well as in academia. Quantitative study on the topic by Dowell *et al.*[2] show that the lifetime (i.e. QE) decreases exponentially as water partial pressure (Torr) at photocathode surfaces increase. In water partial pressure range of $10^{-12}$ to $10^{-9}$ Torr that they reported, the lifetime degreased from 10,000 to 0.1 hours. The pressure range mentioned above is classified as ultrahigh vacuum (UHV) thus it provides an idea as to how sensitive the bialkali photocathodes are to a presence of water molecules. We confirmed that simple calculation based on ideal gas law indicate a trace amount of water molecules in graphene substrates used in this study (below 0.1 atomic %, which is the detection limit of XPS) is enough to reproduce a photocathode environment equivalent to the water partial pressure of $10^{-9}$ Torr. This is an environment that degraded the lifetime of bialkali photocathodes by 5 orders of magnitudes compared to that of $10^{-12}$ Torr thus supports our discussion that residual water molecules in our graphene even after annealing at 350 °C could be enough to induce observed QE decrease.

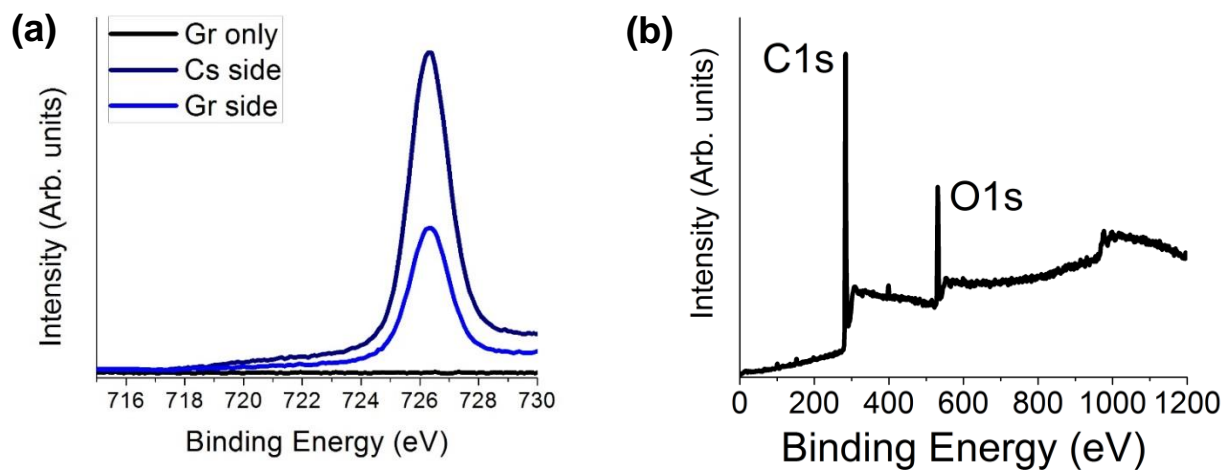

Figure S9 - (a) Narrow scan XPS spectra for Cs 3d 5/2 peak of free-standing graphene substrate on Ni TEM mesh grid after $Cs_3Sb$ deposition compared with a pristine graphene; deposited (dark blue) and non-deposited side (blue), graphene only (black) (b) Wide scan XPS spectrum of free-standing graphene substrate on Ni TEM mesh grid prior to the deposition and annealing.

*S10: Effects of number of graphene layers on photocathode performance*

Work function of $K_2CsSb$ bialkali photocathode interfaced with graphene, quantum tunneling probability of photoelectrons through graphene, and optical absorbance of graphene are the three major factors that affect QE of the photocathodes depending on the number of graphene encapsulation layer. Our DFT calculation shows that work function of $K_2CsSb$ photocathode interfaced with graphene increases from 3.70 to 4.25 eV when the number of graphene is increased from 1 to 2, and a trend continues for the 3 layer case toward that of graphite (~4.5 eV). This is in sharp contrast to the case of Cu metal photocathodes interfaced with graphene. There was no observable difference in the work function depending on the number of graphene layers because the charge transfer between Cu was negligibly small. In the case of $K_2CsSb$ photocathode, however, the charge transfer between interfaced graphene is large as the details are discussed in the main text. Therefore, the work function increases in the range of several hundred meV as the number of graphene layer increase. In order for photoelectrons in $K_2CsSb$ to be emitted into vacuum, they need to tunnel through graphene *via* quantum effect. Tunneling probability of electrons that go through an energy barrier has exponential decay dependence on the thickness. Graphene on $K_2CsSb$ photocathode could be considered as an energy barrier with an atomic thickness. We are currently performing detailed investigations on the topic based on a theoretical model approach. The obtained results, which will be reported elsewhere soon, show that overall tendency of the exponential decay for thicker graphene still holds although there are unique features in the tunneling probability as a function of electron energy. The last component is the 2.3 % per layer optical absorbance of graphene in the visible range. Incident photons that reach $K_2CsSb$ photocathodes decrease by the above mentioned value per additional graphene layer thus directly decreases the QE by the same ratio. The effect of number of graphene layers on the photocathode performance is a combination of above three factors and the details are currently under investigation.

*S11: Applied voltage dependence of photoelectrons*

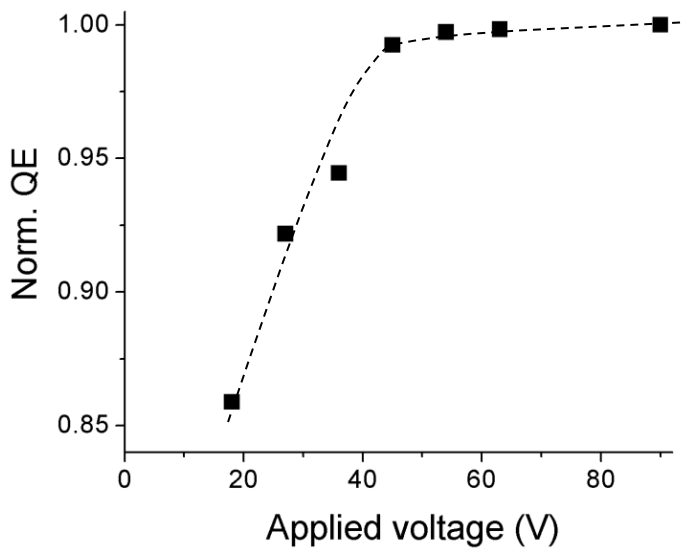

Figure S11 - Applied voltage dependence of photoelectric current from $K_2CsSb$ on free-standing graphene substrates in our experimental setup. Incident light spot size and power were kept constant for all voltages. Trend line is drawn as a guide to the eye.